\documentclass[preprint, showpacs,pra, floatfix]{revtex4}%
\usepackage{graphicx}
\usepackage{dcolumn}
\usepackage{bm}
\usepackage{amsmath}
\usepackage{amsfonts}
\usepackage{amssymb}%
\providecommand{\U}[1]{\protect\rule{.1in}{.1in}}
\begin{document}
\author{R. Cabrera, C. Rangan, W. E. Baylis}
\affiliation{Department of Physics, University of Windsor, Windsor, ON N9B 3P4. Canada}
\title{Overcoming the $su(2^{n})$ sufficient condition for the coherent control of
$n$-qubit systems}

\begin{abstract}
We study quantum systems with even numbers $N$ of levels that are completely
state-controlled by unitary transformations generated by Lie algebras
isomorphic to $sp(N)$ of dimension $N(N+1)/2$. These Lie algebras are smaller
than the respective $su(N)$ with dimension $N^{2}-1$. We show that this
reduction constrains the Hamiltonian to have symmetric energy levels. An
example of such a system is an $n$-qubit system. Using a geometric
representation for the quantum wave function of a finite system, we present an
explicit example that shows a two-qubit system can be controlled by the
elements of the Lie algebra $sp(4)$ (isomorphic to $spin(5)$ and $so\left(
5\right)  $) with dimension ten rather than $su(4)$ with dimension fifteen.
These results enable one to envision more efficient algorithms for the design
of fields for quantum-state engineering, and they provide more insight into
the fundamental structure of quantum control.

\end{abstract}

\maketitle

\section{INTRODUCTION}

The coherent control of an $N$-level quantum system is of interest in fields
such as chemical dynamics \cite{RabitzScience2006}, quantum information
processing~\cite{LloydPRL1999}, and quantum
communication~\cite{RoosScience2004}. It is well
known~\cite{Brockett1972,RamakrishnaPRA1995} that for an $N$-level system to
be completely controllable, it is sufficient that the free-evolution
Hamiltonian, along with the interaction Hamiltonian (which could involve a
sequence of steps) and all possible commutators among them, form a Lie algebra
of dimension $N^{2}$, which in general is taken to be $u(N)$. Recently, it has
been shown~\cite{AlbertiniIEEE2003} that state-to-state controllability can be
achieved with a Lie algebra isomorphic to $sp(N)$ with dimension $N(N+1)/2$.
(We use the notation $sp\left(  N\right)  $ for the algebra of the group
$Sp\left(  N\right)  $ of $N\times N$ unitary symplectic matrices, as for
example in the text by Jones \cite{Jones98}. Other authors denote the same
group by $Usp(N)$ \cite{Gil74} or $Sp(N/2)$ \cite{AlbertiniIEEE2003}.) In this
paper, we show by calculating the Cartan subalgebra that this reduction places
a restriction on the types of systems that can be state-to-state controlled.
Specifically, not only do the systems have to have an even number of energy
levels~\cite{AlbertiniIEEE2003}, their field-free energy levels must be
symmetrically distributed about an average. An example of such a system is a
multi-qubit system that has $N=2^{n}$ energy levels. This result in quantum
control is important both for developing optimal control schemes in quantum
computing \cite{Geremia2004,KhanejaBrockett} and for finding algorithms to
calculate applied fields for quantum-state engineering \cite{Rau2000}.

The control equations can be derived from the time-dependent Schr\"{o}dinger
equation
\begin{equation}
\dot{x}(t)=\left(  A+\sum_{i=1}^{m}u_{i}(t)B_{i}\right)  x(t),
\label{control-equation}%
\end{equation}
where the state vectors $x(t)\in\mathbb{C}^{n}$, give the amplitudes in a
basis of free-evolution eigenstates, $A$ and $B_{i}$ are constant matrices,
and the real scalar functions $u_{i}(t)$ are the control fields. The evolution
of an $N$-level system can be studied by integrating the corresponding matrix
equation in which $x(t)$ is replaced by a matrix $X(t)$, each column of which
represents an independent state; one follows the evolution of $X(t)$ from the
identity matrix $X(0)=I$. If $A$ and $B_{i}$ are anti-Hermitian, the solutions
of $x(t)$ have constant norms $|x(t)|$ and can thus be viewed as lying on a
sphere, and the groups that define the complete controllability of
Eq.~(\ref{control-equation}) for general systems are those summarized in
\cite{Brockett1972}.

In this paper, we study and independently demonstrate a sufficient condition
suggested by Refs.~\cite{Montgomery-Samelson,Brockett1972,AlbertiniIEEE2003}
for establishing controllability of a common class of systems that uses
$sp(N)$ Lie algebras, which are smaller, namely of dimension $N(N+1)/2$,
compared with $N^{2}-1$ for $su(N)$ or $N^{2}$ for $u(N)$. We show that the
Cartan subalgebra of $sp\left(  N\right)  $ restricts its application to
systems where the free-evolution Hamiltonian has a symmetric distribution of
energy levels about an average. These systems are a subset of the general ones
discussed in references \cite{Brockett1972,RamakrishnaPRA1995}. As an example,
we illustrate explicitly that a system with four levels (a two-qubit system)
is controllable with $sp(4)$, which is isomorphic to the $spin(5)$ and $so(5)$
algebras, and which has 10 dimensions and is thus smaller than $su(4)$ with
its 15 dimensions. Similarly, a system with eight levels (a three-qubit
system) is controllable with a Lie algebra of dimension 36, significantly
smaller than $su(8)$ with its 63 dimensions.

\section{SUFFICIENT CONDITION FOR STATE CONTROLLABILITY}

The wave function $\Psi$ is constructed as a unitary transformation of a
reference or pass state~\cite{TuriniciChemPhys2001} represented in a geometric
representation by the primitive projector $P$. The unitary transformation is
an exponential operator of anti-Hermitian elements of the Lie algebra for the
system:
\begin{equation}
\Psi=e^{\mathbf{a}}P,\,\mathbf{a}\in\text{Lie algebra},\, \label{Psi}%
\end{equation}
and $P$ can be represented by the singular matrix%
\[
P=%
\begin{pmatrix}
1 & 0 & \dots\\
0 & 0 & \ldots\\
\vdots & \vdots & \ddots
\end{pmatrix}
\]
with $0s$ everywhere except at the upper-left diagonal position. One can
verify the normalization $\operatorname*{tr}(\Psi^{\dagger}\Psi
)=\operatorname*{tr}(Pe^{-a}e^{a}P)=\operatorname*{tr}(P)=1$. In this form,
the wave function, as an element of a Clifford algebra, represents an
arbitrary single state of the system as a square matrix, corresponding to $X$
mentioned in the previous paragraph but with a single nonvanishing column on
the leftmost side.

Our sufficient condition for a Lie algebra that governs the pure-state control
of a quantum system is based on the following: the parametrization of the wave
function using unitary exponential operators $e^{\mathbf{a}}$ of the Lie
algebra defines a complete control scheme if we are able to reach an arbitrary
ray in the complete state space. We illustrate the procedure first in general
terms and then give explicit examples.

We require that for any pair of basis states $\psi_{j},\psi_{k}$ of the state
space, there exists an anticommuting pair of antihermitian elements
$\mathbf{a}_{kj},\mathbf{b}_{kj}$ of the algebra that relates them:%
\begin{align}
\psi_{k}  &  =\mathbf{a}_{kj}\psi_{j}=-i\mathbf{b}_{kj}\psi_{j}\\
\mathbf{a}_{kj}  &  =-\mathbf{a}_{kj}^{\dag},\ \mathbf{b}_{kj}=-\mathbf{b}%
_{kj}^{\dag},\ \mathbf{a}_{kj}\mathbf{b}_{kj}+\mathbf{b}_{kj}\mathbf{a}%
_{kj}=0.\nonumber
\end{align}
It is important to remember here that the basis states have the projective
form (\ref{Psi}) of a minimal left ideal of the algebra. Assuming unit
normalization $\left(  \mathbf{a}_{kj}\right)  ^{2}=-1=\left(  \mathbf{b}%
_{kj}\right)  ^{2},$ it follows that we can write $\psi_{k}=\exp\left[
\mathbf{a}_{kj}\pi/2\right]  \psi_{j}=-i\exp\left[  \mathbf{b}_{kj}%
\pi/2\right]  \psi_{j},$ and the more general superposition,%
\begin{align}
&  \exp\left[  \left(  \mathbf{a}_{kj}\cos\phi+\mathbf{b}_{kj}\sin\phi\right)
\frac{\theta}{2}\right]  \psi_{j}=\psi_{j}\cos\frac{\theta}{2}+\psi
_{k}e^{i\phi}\sin\frac{\theta}{2}\nonumber\\
&  =\exp\left(  \mathbf{c}_{kj}\frac{\phi}{2}\right)  \exp\left(
\mathbf{a}_{kj}\frac{\theta}{2}\right)  \exp\left(  -\mathbf{c}_{kj}\frac
{\phi}{2}\right)  , \label{transop}%
\end{align}
is expressed as a continuous \textquotedblleft rotation\textquotedblright%
\ with real angle parameters $\theta,\phi,$ in state space, where%
\[
\mathbf{c}_{kj}=\frac{1}{2}\left[  \mathbf{a}_{kj}\mathbf{b}_{kj}%
-\mathbf{b}_{kj}\mathbf{a}_{kj}\right]
\]
is another element of the Lie algebra and we noted that $\mathbf{c}%
_{kj}\mathbf{a}_{kj}=\mathbf{b}_{kj}.$ Products of such unitary operators
allow transitions from one basis state to any linear combination of the
states. One additional element $\mathbf{b}_{jj}$ is needed to simply change
the complex phase of $\psi_{j}:$%
\begin{equation}
i\psi_{j}=\mathbf{b}_{jj}\psi_{j}.
\end{equation}
The elements $\mathbf{a}_{kj},\mathbf{b}_{kj},\mathbf{c}_{kj}$ are generators
of the control group and represent the effect of coupling fields. Given any
initial basis state $\psi_{j},$ a general state of the system is a real linear
combination%
\begin{equation}
\Psi=\sum_{k}\left(  \alpha_{kj}\mathbf{a}_{kj}+\beta_{k}j\mathbf{b}%
_{kj}\right)  \psi_{j},\ \alpha_{kj},\beta_{kj}\in\mathbb{R} \label{Psixpsn}%
\end{equation}
of the $\mathbf{a}_{kj}$ and $\mathbf{b}_{kj}$ generators operating on
$\psi_{j}$, where for notational convenience we write $\mathbf{a}_{jj}=1.$ In
practice, the elements $\mathbf{a}_{kj},\mathbf{b}_{kj},\mathbf{c}_{kj}$ are
members of the same small set. As we demonstrate below, a set of $N$ distinct
elements is sufficient to generate a Lie algebra of $N\left(  N+1\right)  /2$ dimensions.

Calculating the Lie algebra of a higher-dimensional system can require
intensive computations, but there is an elegant and efficient approach using
techniques of Clifford's geometric algebra. The $N$-level quantum system can
be described using multivectors in a geometric algebra. The bivectors are well
known as generators of the spin groups, and it has been shown \cite{Doran93}
that in fact every Lie group can be represented as a spin group. Here we
introduce the possibility of using the full set of anti-Hermitian multivectors
(including, for example, trivectors and six-vectors) to generate the control
group. We illustrate our method with examples of one- and two-qubit systems,
and then generalize to show how the control of an $n$-qubit system can be
achieved by a Lie algebra generally smaller than $su\left(  N\right)  .$

\subsection{Example: Single Qubit Control}

In the simplest example, Clifford's geometric algebra $C\!\ell_{3}$ of
three-dimensional Euclidean space enables us to describe a single qubit
($N=2$) \cite{Bay03a}. In this case, Pauli spin matrices can represent the
three orthonormal vectors (basis elements of grade~$1$): $\mathbf{e}%
_{j}=\boldsymbol{\sigma}_{j},~j=1,2,3.$ The products of grade $2$
\begin{equation}
\boldsymbol{\ }\mathbf{e}_{12}=\mathbf{e}_{1}\mathbf{e}_{2},\ \mathbf{e}%
_{23}=\mathbf{e}_{2}\mathbf{e}_{3},\ \mathbf{e}_{31}=\mathbf{e}_{3}%
\mathbf{e}_{1}~,
\end{equation}
form a basis for the bivector space and generate rotations. There is a single
independent element of grade $3$, namely the trivector
\begin{equation}
\mathbf{e}_{123}=\mathbf{e}_{1}\mathbf{e}_{2}\mathbf{e}_{3}\boldsymbol{~},
\end{equation}
whose matrix representation is $i$ times the unit matrix. These elements along
with the identity span the full linear space of the closed algebra
$C\!\ell_{3}$.

We can take the basis states of the system to be $\psi_{0}=P$ and $\psi
_{1}=\mathbf{e}_{13}P.$ Then we note by the \textquotedblleft
pacwoman\textquotedblright\ property of projectors, \cite{Bay03a,Bay99} namely
$\mathbf{e}_{3}P=P,$ that $i\psi_{1}=i\mathbf{e}_{1}P=\mathbf{e}_{23}P$ and
$i\psi_{0}=\mathbf{e}_{12}P.$ The $N=2$ generators $\mathbf{a}_{10}%
=\mathbf{e}_{13},\mathbf{\ c}_{10}=-\mathbf{e}_{12},$ generate the control Lie
algebra $spin\left(  3\right)  ,$ which is isomorphic to $su\left(  2\right)
,$ $so\left(  3\right)  ,$ and $sp\left(  2\right)  $. An arbitrary state can
be expressed by%
\[
\Psi=\exp\left(  -\mathbf{e}_{12}\frac{\phi}{2}\right)  \exp\left(
\mathbf{e}_{13}\frac{\theta}{2}\right)  \exp\left(  -\mathbf{e}_{12}\frac
{\chi}{2}\right)  P,
\]
which, in fact, is just the Euler-angle expression for the Bloch-sphere
representation the state~\cite{Bay03a}. Note that since the exponents form a
closed Lie algebra, no generators outside of the algebra arise from an
expansion of the unitary operator~\cite{Rau2000}.

We assume a basis for the system in which the free-evolution Hamiltonian
$H_{0}$ is diagonal. Since commutators (Lie products) of $H_{0}$ with the
control transformations are to remain within the Lie algebra, we need to
construct $H_{0}$ from the unit matrix plus elements of the Lie algebra. To
ensure that $H_{0}$ is diagonal, its contributions from the Lie algebra are
restricted to the Cartan subalgebra, defined as the largest set of commuting
generators of the Lie algebra. For the two-state system, the Cartan subalgebra
of $su\left(  2\right)  $ comprises a single element, namely the generator
$\mathbf{e}_{12}=\sigma_{1}\sigma_{2}=i\sigma_{3}$. We thus construct a
general Hamiltonian for a two-level system (apart from an offset energy
proportional to the unit matrix) as
\begin{equation}
H_{0}=-i\mathbf{e}_{12}\omega=\omega%
\begin{pmatrix}
1 & 0\\
0 & -1
\end{pmatrix}
.
\end{equation}

\section{GEOMETRIC REPRESENTATION OF MULTI-QUBIT CONTROL}

For systems of multiple qubits, the orthogonal unit vectors of the appropriate
Clifford algebra can be represented as tensor products (Kronecker products) of
the Pauli matrices as shown in Table \ref{reps}. Bivectors, trivectors, etc.,
can be obtained by the product of the unit orthogonal vectors among themselves.

\begin{center}%
\begin{table}[tbp] \centering
\begin{tabular}
[c]{cc}%
$4D$ and $5D$ & $7D$\\%
\begin{tabular}
[c]{cr}\hline
$\mathbf{e}_{1}=$ & $\boldsymbol{\ \sigma}_{3}\otimes\boldsymbol{\ \sigma}%
_{1}$\\
$\mathbf{e}_{2}=$ & $\boldsymbol{\ \sigma}_{3}\otimes\boldsymbol{\ \sigma}%
_{2}$\\
$\mathbf{e}_{3}=$ & $\boldsymbol{\ \sigma}_{3}\otimes\boldsymbol{\ \sigma}%
_{3}$\\
$\mathbf{e}_{4}=$ & $-\boldsymbol{\ \sigma}_{2}\otimes1$\\
$\mathbf{e}_{5}=$ & $-\boldsymbol{\ \sigma}_{1}\otimes1$\\\hline
\end{tabular}
&
\begin{tabular}
[c]{crcr}\hline
$\mathbf{e}_{1}=$ & $1\otimes\boldsymbol{\ \sigma}_{3}\otimes
\boldsymbol{\ \sigma}_{1}$ & $\,\mathbf{e}_{6}=$ & $\boldsymbol{\ \sigma}%
_{1}\otimes\boldsymbol{\ \sigma}_{1}\otimes1$\\
$\mathbf{e}_{2}=$ & $1\otimes\boldsymbol{\ \sigma}_{3}\otimes
\boldsymbol{\ \sigma}_{2}$ & $\,\mathbf{e}_{7}=$ & $\boldsymbol{\ \sigma}%
_{2}\otimes\boldsymbol{\ \sigma}_{1}\otimes1$\\
$\mathbf{e}_{3}=$ & $1\otimes\boldsymbol{\ \sigma}_{3}\otimes
\boldsymbol{\ \sigma}_{3}$ &  & \\
$\mathbf{e}_{4}=$ & $1\otimes\boldsymbol{\ \sigma}_{2}\otimes1$ &  & \\
$\mathbf{e}_{5}=$ & $\boldsymbol{\ \sigma}_{3}\otimes\boldsymbol{\ \sigma}%
_{1}\otimes1$ &  & \\\hline
\end{tabular}
\end{tabular}
\caption{A matrix representation of orthonormal vectors for some dimensions.
The $4\times4$ matrix representation for $5D$ is not faithful for the
universal Clifford algebra $C\!\ell_{5}$ (it is a homomorphism rather than an
isomorphism) but does represent all bivectors uniquely and is therefore
adequate for state control.}\label{reps}%
\end{table}%

\end{center}

Any homogeneous multivector (comprising elements of a single grade $g$) in the
real Clifford algebra $C\!\ell_{n}$ for an $n$-dimensional Euclidean space can
be classified as Hermitian or anti-Hermitian according to its grade. Elements
of grade 0,1,4,5,8,9 or generally whenever the grade is $0$ or $1\,\,mod\,\,4$%
, are Hermitian whereas those of other grades are anti-Hermitian. This is
important because the bivectors on one hand, as well as the complete set of
anti-Hermitian multivectors on the other hand, form Lie algebras of compact
groups. In some algebras for Euclidean spaces of odd dimension, as for example
in $C\!\ell_{3}$ or $C\!\ell_{7},$ the highest-grade multivector (the volume
element) is anti-Hermitian but commutes with every element of the algebra, and
it therefore must be excluded from the set of all anti-Hermitian elements that
generates the Lie algebra.

\subsection{Example: Two-Qubit Control}

The four-level system, understood as comprising two qubits, is controlled
using the bivectors plus trivectors of the Clifford algebra $C\!\ell_{4}$ of
four-dimensional Euclidean space or, equivalently, by the bivectors of a
Clifford algebra for a five-dimensional Euclidean space, namely the
nonuniversal Clifford algebra $C\!\ell_{5}\left(  1+\mathbf{e}_{12345}\right)
/2$, a left ideal of $C\!\ell_{5},$ which is isomorphic to $C\!\ell_{4}.$
These bivectors generate the $spin(5)$ algebra, which is isomorphic to
$so\left(  5\right)  $ and to $sp(4)$. The projector for two-qubits, can be
represented in terms of bivectors $\mathbf{e}_{jk}$ (see Table \ref{reps}) by
\begin{equation}
P=\frac{1}{4}(1-i\mathbf{e}_{12})(1+i\mathbf{e}_{45}).
\end{equation}
The dimension of the control algebra is ten. Because the elements form a
closed algebra, in this case $spin\left(  5\right)  ,$ we know that no other
generators are needed for state control. The Cartan subalgebra in this case is
two dimensional so that there are two diagonal generators among the $spin(5)$
generators, from which we can construct the free-evolution Hamiltonian (apart
from a constant offset and with $\hbar=1$)
\begin{equation}
H_{0}=\frac{i}{2}(\omega_{2}+\omega_{1})\mathbf{e}_{45}-\frac{i}{2}(\omega
_{2}-\omega_{1})\mathbf{e}_{12}~.
\end{equation}
This Hamiltonian has symmetric eigenenergies as represented in figure
(\ref{Fig1sym4levBW})
\begin{equation}
H_{0}=%
\begin{pmatrix}
\omega_{2} & 0 & 0 & 0\\
0 & \omega_{1} & 0 & 0\\
0 & 0 & -\omega_{1} & 0\\
0 & 0 & 0 & -\omega_{2}%
\end{pmatrix}
.
\end{equation}
The sufficient condition for state controllability thus leads to a class of
systems \emph{with energy levels symmetrically distributed about a center,}
such as those that can be found in trapped-ion qubits \cite{RanganPRL2004} or
coupled spins~\cite{ChuangNature2001}.

The unitary transition operators among the eigenstates can be expressed [see
Eq. (\ref{transop})] in the form $\exp\left(  \mathbf{c}\phi/2\right)
\exp\left(  \mathbf{a}\theta/2\right)  \exp\left(  -\mathbf{c}\phi/2\right)
,$ where $\theta$ determines the magnitudes of the state amplitudes and $\phi$
gives the relative phase. The transition between states is complete when
$\theta=\pi,$ as in a $\pi$ pulse. Table \ref{Gentrans} shows the generators
$\mathbf{a,c}$ for each transition in the 2-qubit system. Note that with
$\theta=\pm\pi/2,$ the partial transitions $1\leftrightarrow2,0\leftrightarrow
3$ induced by the coupled-qubit bivector $\mathbf{e}_{24},$ create the four
entangled Bell states.

\begin{center}%
\begin{table}[tbp] \centering
\begin{tabular}
[c]{|l|l|l|}\hline
$\mathbf{a}$ & $\mathbf{c}$ & \textbf{Transitions}\\\hline
$0$ & $\mathbf{e}_{12}$ & $0\leftrightarrow0,1\leftrightarrow
1,2\leftrightarrow2,3\leftrightarrow3,$\\
$\mathbf{e}_{13}$ & $\mathbf{e}_{12}$ & $0\leftrightarrow1,2\leftrightarrow
3$\\
$\mathbf{e}_{24}$ & $\mathbf{e}_{12}$ & $1\leftrightarrow2,0\leftrightarrow
3$\\
$\mathbf{e}_{35}$ & $\mathbf{e}_{45}$ & $0\leftrightarrow2,1\leftrightarrow
3$\\\hline
\end{tabular}
\caption{Generators for transition operators in 2-qubit systems (see text).}\label{Gentrans}%
\end{table}%

\end{center}

Thus all the transitions, together with control of the relative phase, require
no more than the five nonzero elements in Table~\ref{Gentrans} and commutators
of these elements give all ten independent elements of $spin\left(  5\right)
.$ However, only four of the five are required in a minimal set, since for
example $\mathbf{e}_{45}$ can be obtained from the other four:%
\begin{align*}
\frac{1}{2}\left[  \mathbf{e}_{12},\mathbf{e}_{24}\right]   &  =\mathbf{e}%
_{14}\\
\frac{1}{2}\left[  \mathbf{e}_{13},\mathbf{e}_{35}\right]   &  =\mathbf{e}%
_{15}\\
\frac{1}{2}\left[  \mathbf{e}_{15},\mathbf{e}_{14}\right]   &  =\mathbf{e}%
_{45}=\exp\left(  \mathbf{e}_{15}\pi/4\right)  \mathbf{e}_{14}\exp\left(
-\mathbf{e}_{15}\pi/4\right)  .
\end{align*}
Fewer than four is easily seen to be insufficient to generate all the elements
of $spin\left(  5\right)  ,$ so that four is the number of elements that is
necessary and sufficient for state control of an arbitrary 2-qubit system. The
anti-Hermitian multivectors used to define controllable schemes are summarized
in Table \ref{contr:schemes} for small systems.

\begin{center}%
\begin{table}[tbp] \centering
\begin{tabular}
[c]{|l|c|c|c|c|}\hline
Clifford Algebra & qubits & $N$ & Lie algebra & Dim\\\hline
$C\!\ell_{3}$ Bivectors\thinspace\thinspace\ only & $1$ & $2$ & $su(2)$ &
$3$\\
$C\!\ell_{4}$ Anti-Hermitian & $2$ & $4$ & $sp(4)$ & $10$\\
$C\!\ell_{5}$ Bivectors\thinspace\thinspace\ only & $2$ & $4$ & $spin(5)\cong
sp\left(  4\right)  $ & $10$\\
$C\!\ell_{6}$ Anti-Hermitian & $3$ & $8$ & $sp(8)$ & $36$\\\hline
\end{tabular}
\caption{Lie algebras and their controllable $n$-qubit systems. $N=2^n$ is the number of levels and also the minimum number of elements needed to generate the entire algebra.}\label{contr:schemes}%
\end{table}%

\end{center}

\begin{figure}
[ptb]
\begin{center}
\includegraphics[
height=1.5134in,
width=5.6005in
]%
{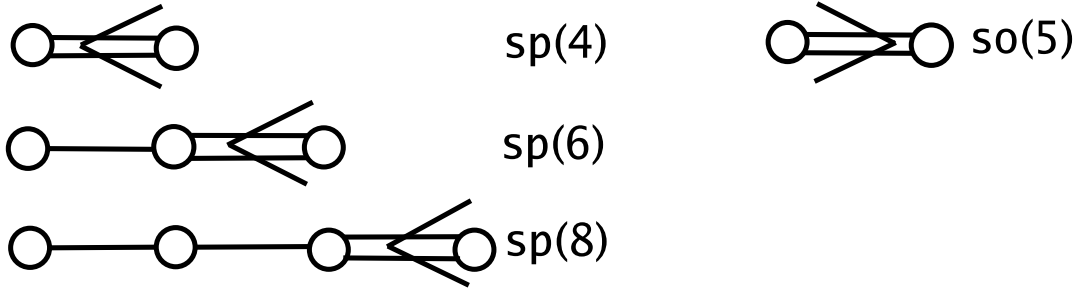}%
\caption{Dynkin diagrams corresponding to some of the Symplectic Lie algebras
in low dimensions and the case for $so(5),$ which is isomorphic to
$spin\left(  5\right)  .$}%
\label{Fig2DynkinGraph}%
\end{center}
\end{figure}

\begin{figure}
[ptb]
\begin{center}
\includegraphics[
height=3.2655in,
width=2.495in
]%
{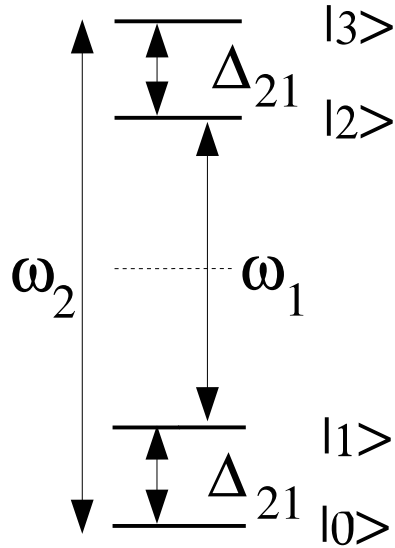}%
\caption{Symmetric energy levels of a 2-qubit system with two interaction
fields.}%
\label{Fig1sym4levBW}%
\end{center}
\end{figure}

The Lie algebras of interest are of dimension $N(N+1)/2$, which is the same as
the dimension of the symplectic Lie algebras $sp(N)$ (for even $N$). The
Dynkin diagrams for lower dimension are shown in figure \ref{Fig2DynkinGraph}
including the case of $sp(4)$ to show the isomorphism with $so(5)$ and thus
with $spin\left(  5\right)  .$%

\section{EXPLICIT CONTROL SCHEME}

The method is readily extended to higher even values of $N.$ An explicit
control scheme can show that an arbitrary superposition in a quantum system
with even number of energy levels that are symmetrically distributed about an
offset can be produced from another arbitrary superposition using a set of
fields, and the Lie algebras generated by these field-couplings are of
dimension $N(N+1)/2$. This scheme is based on the subspace controllability
theorem \cite{bloch-brockett-rangan-2006} that describes the method of
transferring any superposition of states to any other superposition through a
pivot state (pass state). This builds on the work done by Eberly and coworkers
on the control of harmonic oscillator states~\cite{Law-Eberly,Kneer-Law}.

In the general case of the even $N$-level system with symmetric energies, this
scheme is implemented by transferring population in any superposition of
states to the ground state $|0\rangle$ through a sequential application of
fields. (In Table \ref{Gentrans}, we show the fields connecting all energy
states, and in practise some of these these may correspond to qubit-qubit
couplings. However, this scheme will succeed with any sequentially connected
quantum transfer graph~\cite{RanganJMP2005}) To obtain any arbitrary
final-state superposition, the time-reversed sequence of fields is applied
starting from $|0\rangle$. Since the system is finite, we conclude that it is
arbitrarily controllable. Note that $n$-qubit systems are all cases of the
general even-level system with symmetric energy distributions.%

The control algebra for this scheme contains only $N(N+1)/2$ elements, which
can be always constructed defining an initial set of $N$ generators with
representation matrices of the form%
\begin{equation}%
\begin{array}
[c]{ll}%
\begin{pmatrix}
0 & 1 & 0 & \ldots & 0 & 0 & 0\\
-1 & 0 & 0 &  & 0 & 0 & 0\\
0 & 0 & 0 &  & 0 & 0 & 0\\
\vdots &  &  &  &  &  & \\
0 & 0 & 0 &  & 0 & 0 & 0\\
0 & 0 & 0 &  & 0 & 0 & 1\\
0 & 0 & 0 & \ldots & 0 & -1 & 0
\end{pmatrix}
&
\begin{pmatrix}
0 & 0 & 0 & \ldots & 0 & 0 & 0\\
0 & 0 & 1 &  & 0 & 0 & 0\\
0 & -1 & 0 &  & 0 & 0 & 0\\
\vdots &  &  &  &  &  & \\
0 & 0 & 0 &  & 0 & 1 & 0\\
0 & 0 & 0 &  & -1 & 0 & 0\\
0 & 0 & 0 & \ldots & 0 & 0 & 0
\end{pmatrix}
\ldots\\%
\begin{pmatrix}
0 & i & 0 & \ldots & 0 & 0 & 0\\
i & 0 & 0 &  & 0 & 0 & 0\\
0 & 0 & 0 &  & 0 & 0 & 0\\
\vdots &  &  &  &  &  & \\
0 & 0 & 0 &  & 0 & 0 & 0\\
0 & 0 & 0 &  & 0 & 0 & i\\
0 & 0 & 0 & \ldots & 0 & i & 0
\end{pmatrix}
&
\begin{pmatrix}
0 & 0 & 0 & \ldots & 0 & 0 & 0\\
0 & 0 & i &  & 0 & 0 & 0\\
0 & i & 0 &  & 0 & 0 & 0\\
\vdots &  &  &  &  &  & \\
0 & 0 & 0 &  & 0 & i & 0\\
0 & 0 & 0 &  & i & 0 & 0\\
0 & 0 & 0 & \ldots & 0 & 0 & 0
\end{pmatrix}
\ldots.
\end{array}
\end{equation}
For example, for two qubits, this initial set of generators is equivalent to
\begin{equation}
\{-\mathbf{e}_{13},-(\mathbf{e}_{15}+\mathbf{e}_{24})/2,\mathbf{e}%
_{23},(\mathbf{e}_{14}-\mathbf{e}_{25})/2\}.
\end{equation}
The complete algebra is then found from all the new possible independent
commutators calculated recursively until the linear space is
exhausted~\cite{RanganJMP2005}. From the complete algebra, only the Cartan
subalgebra with $N/2$ elements can be used to define the field-free
Hamiltonian of the system. As a result, the field-free Hamiltonian cannot have
an arbitrary distribution of energy levels but must have the energy levels
symmetrically distributed around an average (the offset) energy.

\section{SUMMARY}

A Lie algebra of $N(N+1)/2$ elements---significantly fewer than $N^{2}$ ---is
shown to be sufficient for arbitrary control of an even-level quantum system
with symmetric energy levels, specifically of $n=log_{2}(N)$-qubit systems.
All the elements of the algebra can be generated from a minimal set of $N$
elements, which is the minimum number of generators for state control of the
$N$-level system. These results have the potential to lead to more efficient
optimal-control schemes for quantum state engineering and production of
entangled states.

\section*{ACKNOWLEDGEMENTS}

Our research is supported by the Natural Sciences and Engineering Research
Council of Canada.

\end{document}